\documentclass[showpacs,preprintnumbers,amsmath,amssymb]{revtex4}

\date{}

\usepackage{graphicx}
\usepackage{dcolumn}
\usepackage{bm}

\begin{document}

\title{\bf Melting curve of face-centred-cubic nickel from first principles calculations }

\author{Monica Pozzo$^1$}
\author{Dario Alf\`{e}$^1$}%
\email{d.alfe@ucl.ac.uk}

\affiliation{
  $^1$Department of Earth Sciences, Department of Physics and
  Astronomy, London Centre for Nanotechnology and Thomas Young Centre@UCL, University
  College London, Gower Street, London WC1E 6BT, United Kingdom }


\begin{abstract}
The melting curve of Ni up to 100 GPa has been calculated using first principles methods based on density functional theory (DFT). 
We used two complementary approaches: i) coexistence simulations with a reference system and then free energy corrections between DFT and the reference system, and
ii) direct DFT coexistence using simulation cells including 1000 atoms. The calculated zero pressure melting temperature is slightly underestimated at $1637 \pm 10$ K 
(experimental value is 1728 K),
and at high pressure is significantly higher than recent measurements in diamond anvil cell experiments  [Phys. Rev. B {\bf 87}, 054108 (2013)]. The zero pressure DFT 
melting slope is calculated to be $30 \pm 2$ K, in good agreement with the experimental value of 28 K.
 \end{abstract}

\maketitle

\section{Introduction}

The melting curves of transition metals have recently attracted significant interest, particularly because of long-standing controversies mainly due to the different 
predictions of high pressure melting temperatures between diamond anvil cell (DAC) experiments (e.g. ~\cite{boehler93,errandonea01}) and shock-wave 
(SW) experiments (e.g. ~\cite{brown86, hixson89, nguyen04, dai09}). Although the range of pressure
explored by the two techniques is quite different, and the DAC data need to be extrapolated before they can be directly compared to SW data at the same pressure, 
it is clear that large differences appear in the predicted high pressure melting temperatures of several transition metals, particularly 
molybdenum~\cite{hixson89, errandonea01,santamaria09} and tantalum~\cite{errandonea01,dai09}.  
Recent DAC work, in which different diagnostics used to identify the melting transition  
are proposed, seems to reconcile some of these differences, at least for tantalum~\cite{dewaele10} and iron~\cite{jackson13,anzellini13}.

On the theoretical side a large number of attempts at calculating the melting curves of several transition metals have been proposed, although most of these studies were
based on empirical interatomic potentials (e.g. ~\cite{laio00,belonoshko00,weingarten11,luo10,demaske13}). 
Different calculations for the same material can also show significant discrepancies, which are due to the different quality of the 
respective interatomic potentials. On the other hand, calculations based on first principles approaches, mainly within density functional theory (DFT), have been shown to 
be very accurate at predicting the correct pressure behaviour of the melting temperature of several materials, including the transition metals tantalum~\cite{taioli07} 
and iron~\cite{alfe02,alfe09}.

Here we report on the DFT calculation of the melting curve of nickel from 0 to 100 GPa. Our results are  higher than recent DAC measurements~\cite{errandonea13}, and 
lower than previously reported calculations based on empirical potentials~\cite{luo10,weingarten11}, and in agreement with that computed by Ko\v{c}i et al.~\cite{koci06},
who also used an approach based on empirical potentials.   

The paper is organized as follows. Sec. II contains the techniques used in the calculations. Results and discussion are presented in Sec. III. Conclusions and final remarks follow in Sec. IV.

\section{Methods}
\begin{figure}[htbp]
   \centering
   \includegraphics[width=4in]{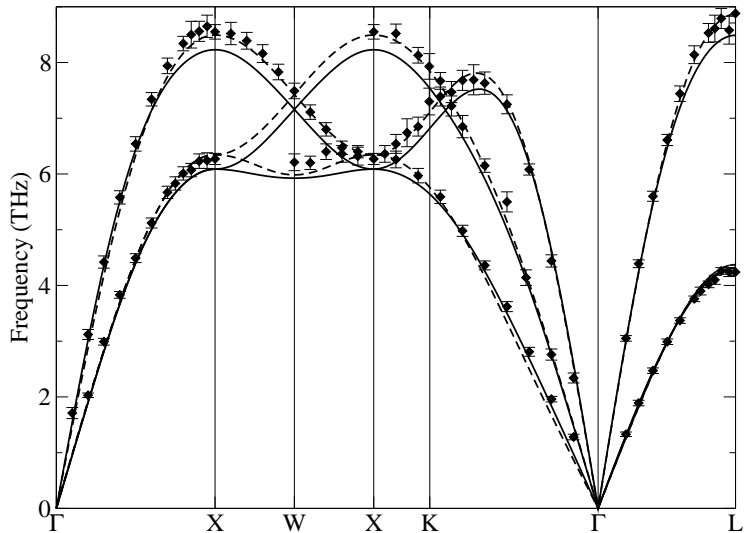} 
   \caption{Phonon dispersions of Ni. Solid lines: present calculations at the DFT-PBE equilibrium volume at T= 296 K  ($a_0 = 3.539$ \AA); dashed lines: 
   present calculations at the DFT-PBE 
   classical equilibrium volume at T = 0 K ($a_0 = 3.52$ \AA); diamonds with error bars:
    experimental data from Ref.~\protect\cite{birgeneau64}.}.
   \label{fig:phon}
\end{figure}

The calculations are based on DFT in the finite temperature formulation
due to Mermin~\cite{mermin65},  with the exchange-correlation potential known as PBE~\cite{perdew96}, as implemented in the {\sc vasp} code~\cite{kresse96}. 
We used the projector-augmented-wave (PAW) formalism~\cite{blochl94,kresse99}, and for the majority of the calculations a nickel PAW potential with an [Ar] core 
(10 electrons in valence) and an outmost cutoff radius of 1.21 \AA.  
Single particle wave functions were expanded in plane-waves (PW), with exact details of the PW cutoff reported below. 
With this potential the lattice parameter of the face-centred-cubic crystal of Ni at 
zero temperature (and no zero point motion) is calculated to be 3.52 \AA, the bulk modulus  is 194 GPa, and the magnetic moment is 0.62 $\mu_B$/atom. To compare with the 
experimental values at T = 296 K we have calculated the free energy of the crystal 
 in the quasi-harmonic approximation, using the small displacement method as implemented in the 
{\sc phon} code~\cite{alfe09b}. We used a $4\times4\times4$ supercell (64 atoms) and a displacement of 0.04 \AA. We performed spin-polarised 
calculations using a $ 3\times 3\times 3$ grid of {\bf k}-points to sample the Brillouin zone, and free energies were obtained by integrating phonon frequencies
over a $8\times 8\times 8$ grid of {\bf q}-points (in fact, even a $4\times 4\times 4$ grid of {\bf q}-points would give free energies converged to better than 0.02 meV/atom at
300 K). 
At T = 296 K the lattice parameter, 
bulk modulus and magnetic moments are calculated to be 3.539 \AA, 182 GPa and 0.62 $\mu_B$/atom, being slightly larger, lower and in good agreement with the 
the experimental values of 3.52 \AA, 186 GPa and 0.62 $\mu_B$/atom, respectively. The slight overestimation of the lattice parameter is typical of DFT-PBE, and also agrees
with previously reported results~\cite{dalcorso13}. The volumetric thermal expansion is calculated to be $38 \times 10^{-6} $K$^{-1}$ , in good agreement 
with the experimental value of $39 \times 10^{-6} $K$^{-1}$.
Phonon dispersions at zero pressure are compared with experiments in Fig.~\ref{fig:phon},  where we show calculations both at the classical equilibrium lattice 
parameter at zero temperature, which is the same as the experimental lattice parameter (3.52 \AA ), and at the calculated lattice parameter at T = 296 K (3.539 \AA ). 
The agreement with the experiments is better at the experimental lattice parameter.
Phonons dispersions of Ni have previously been computed by Dal Corso~\cite{dalcorso13}, who obtained similar results. 

The main strategy used here to calculate the melting curve follows the method developed in Ref.~\cite{alfe02b}. The idea is to use a reference potential to compute the melting
curve first, and then correct it by calculating free energy differences between the ab-initio and the reference potentials. We outline the strategy in the following.
  
The thermodynamic condition that determines the melting point $T_m$ at pressure $p$ is given by 
$G^{ls}(p,T_m) = G^l(p,T_m) - G^s(p,T_m)$,
where $G^l(p,T_m)$ and $G^s(p,T_m)$ are the Gibbs free energies of the system in the liquid and the solid 
state at  $p$ and $T_m$. 
In the vicinity of $T_m$
the temperature dependences of $G^l$ and $G^s$ are linear, with the slopes given by the entropies  $S^l$ 
and $S^s$. It is 
straightforward to see that, as the potential energy function is changed from that of the reference to that of the ab-initio potential,  
a small relative shift of the free energy of the liquid with respect to the free energy of the solid,   
$\Delta G^{ls}(T_m)$, where $\Delta$ indicates difference between ab-initio and reference systems, causes a shift in the melting point given by:
\begin{equation}\label{eqn:shift}
 \Delta T_m = \frac{\Delta G^{ls}(T_m)}{S^{ls}}
\end{equation}
 where $S^{ls} = S^l - S^s$ is the entropy change on melting of the reference system.
This relation is exact if the Gibbs free energies are exactly linear, and it is only an approximation otherwise, in which case higher order corrections can
also be calculated (see Ref.~\cite{alfe02b}). 
This argument provides a possible route to the calculation of the ab-initio melting curve: i) we construct a reference system in a way that
its  free energy is as close as possible to the free energy of the DFT system; ii) we compute the melting temperature of the reference system using
standard methods (see below); iii) we compute the difference between the DFT and the reference system melting temperature using the 
linear relation outlined above. 

The Gibbs free energy difference between the DFT and the reference system can be calculated using 
thermodynamic integration~\cite{frenkel96}.
If this difference is small, it can be calculated as~\cite{alfe02b}:
\begin{equation}\label{eqn:dg}
\Delta G = \langle \Delta U \rangle_{ref} - \frac{1}{2 k_{\rm B}T} \langle \delta \Delta U^2\rangle_{ref} + \dots,
\end{equation}
where $k_{\rm B}$ is the Boltzmann constant, $\Delta U = U - U_{ref}$ is the difference between the DFT and the reference 
potential energy functions $U$ and $U_{ref}$, 
respectively, $\delta \Delta U = \Delta U - \langle \Delta U \rangle_{ref}$, and $\langle \cdot \rangle_{ref}$ represents average 
evaluated in the reference isothermal-isobaric ensemble. A similar relation holds for the Helmholtz free energy difference $\Delta F$, which
is obtained by replacing the isothermal-isobaric with the isothermal-isochoric ensemble.  The relation between $\Delta G$ and $\Delta F$ is 
readily shown to be~\cite{alfe02b}:
\begin{equation}\label{eqn:dg1}
\Delta G = \Delta F - \frac{1}{2}\frac{V}{K_T}\Delta p^2,
\end{equation}
where $V$ is the volume, $K_T$ the isothermal bulk modulus and $\Delta p$ the pressure difference between the two systems at volume $V$ 
and temperature $T$. Eqns.~(\ref{eqn:shift}), (\ref{eqn:dg}), (\ref{eqn:dg1}), suggest a strategy to construct the reference potential. We look 
for a reference system for which $\delta \Delta U$ are as small as possible, so we can use Eqn.~(\ref{eqn:dg}) to compute $\Delta G$. Since
we prefer to work with isothermal-isochoric ensemble, we also require $\Delta p$ to be small, so we can compute $\Delta G$ from 
$\Delta F$ using Eqn.~(\ref{eqn:dg1}). Finally, a crucial point in the scheme, we
want to use the linear relation in Eqn.~(\ref{eqn:shift}) to compute the shift of melting temperature between the ab-initio and the reference systems.
We therefore require that the relative shift of free energies between liquid and solid, $\Delta^{ls}$, are as small as possible.   

We chose as reference potential an embedded atom model (EAM)~\cite{daw83} with the form proposed by Sutton
and Chen~\cite{sutton90}. For this model the total energy of the system is written as:
\begin{equation}
E_{tot} = \frac{1}{2}\sum_{i,j \ne i} \epsilon \left ( \frac{a}{r_{ij}} \right )^n  -\epsilon C \sum_i \left ( \sum_{j\ne i} \left ( \frac{a}{r_{ij}} \right )^m \right )^{1/2},
\end{equation}
where $\epsilon, a, C, n, m$ are fitting parameters, and $r_{ij}$ is the distance between two atoms at positions ${\bf r}_i$ and ${\bf r}_j$.
Since the potential has infinite range a cutoff in real space needs to be applied, which we choose to be 6 \AA. The potential was then cut and shifted, in order to eliminate
discontinuities in the energy at the cutoff distance. The remaining discontinuities in the forces are sufficiently small that they do not present any problems in the 
molecular dynamics simulations. 
   
\section{Results and discussion}
To determine the parameters of the potential we initially performed two long DFT simulations, one for the solid and one for the liquid, at 
a pressure close to 30 GPa. From these simulations
we extracted 100 statistically independent configurations. We used these configurations, together with the DFT 
energies and pressures, to fit the parameters of the EAM 
by minimising both $\delta \Delta U^2$ and $\Delta p^2$. The parameters that we obtained were $\epsilon = 3.1774\times 10^{-2}$ eV, 
$a = 3.1323$~\AA, $C=33.5741, n = 8.975, m=3.631$, and we denote as EAM1 the embedded atom model with these parameters. 
Note that the value of $n$ is very close to 9, the value originally chosen by
Sutton and Chen~\cite{sutton90}. This parameter determines the shape of the repulsive part of the potential, which is the main responsible 
for the fluctuations in the total energy as the atoms move around sampling the phase space. 

\begin{figure}[htbp]
   \centering
   \includegraphics[width=4in]{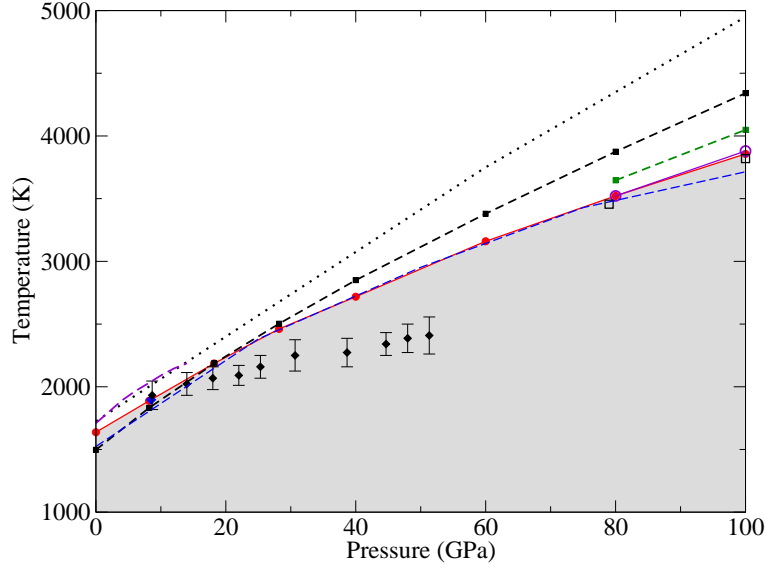} 
   \caption{Melting curve of Ni. Black squares: melting curve of EAM1; green squares: melting curve of EAM2; solid red and open purple circles: 
   DFT-PBE melting curve obtained by correcting the melting curves of EAM1 and EAM2, respectively; dotted lines: EAM melting curve of Ref.~\protect\cite{luo10}; 
   blu dashed line: melting curve of Ref.~\cite{koci06}; violet dashed line: melting curve of Ref.~\cite{weingarten11}; black filled diamonds: 
   DAC experimental values of Ref.~\protect\cite{errandonea13};  open squares (at 79 and 100 GPa): SW based experimental values of Ref.~\protect\cite{urlin66}.}
   \label{fig:melt}
\end{figure}

To compute the melting curve of EAM1 we used the coexistence method. The simulations were performed with cells containing 8000 atoms
($10\times10\times20$), and the constant stress algorithm (NpH) described by Hern\'andez~\cite{hernandez01}. 
We found that the simulation cell retains its zero temperature rectangular shape almost exactly, with only $\simeq 2\%$ strain in the 
direction perpendicular to the solid-liquid interface. 
Simulations were carried out using a 
time step of 3 fs for a total length of 300 ps. The error on the melting temperature was calculated by standard re-blocking procedure~\cite{allen87} and
was found to be less than 5 K. 
We tested size effects by performing simulations on both 1000 and 27000 atoms, 
which showed results essentially identical to those obtained with the 8000-atom simulation cells. 
We also performed simulations at constant volume (NVE), both with the 1000-atom and the 8000-atom simulation cells, and found 
that the resulting melting temperature was indistinguishable from that calculated by using the NpH ensemble. The NVE simulations were performed both with and without 
allowing for the $\simeq 2\%$ strain in the direction perpendicular to the solid-liquid interface, and the effect of the strain was undetectable within error bars of 5 K.
The melting curve of EAM1 is  shown in Fig.~\ref{fig:melt} and reported in Table~\ref{tab:I}.

\begin{table}
\begin{tabular}{ccccccccccccccc}
\hline
$p$ & 0 & 8.2 & 18.2 & 28.2 & 40 & 60 & 80 & 100 \\
$T_m^{\rm EAM1}$ & 1497 & 1832 & 2186 & 2503 & 2850 & 3380 & 3875 & 4341 \\
$\langle \Delta U^{ls}\rangle_{\rm EAM1}$ & 15.3 (0.3) & 8.6 (0.4) & 3.2 (0.7) & -1.5 (0.5) & -6.8 (0.8) & -14.5 (0.7) & -20.7 (0.8) & -29.8 (1.0) \\  
$ -\frac{1}{2k_{\rm B}T}\langle \delta \Delta U_s^2\rangle_{\rm EAM1}$ & -2.4 (0.5) & -2.5 (0.4) & -1.1 (0.4) & -2.3 (0.4) & -2.0 (0.5) & -4.6 (0.7) & -5.8 (1.0) & -7.8 (1.4) \\
$ -\frac{1}{2k_{\rm B}T}\langle \delta \Delta U_l^2\rangle_{\rm EAM1}$  & -3.1 (0.6) & -5.2 (1.3)  & -4.1 (1.8) & -4.6 (1.2) & -6.5 (1.5) & -7.9 (1.3) & -12.5 (2.3) & -15.4 (2.5) \\
$K_T^s$ & 86.8 & 129.3 & 181.0 & 241.6 & 291.6 & 401.4 & 540.3 & 543.6 \\
$K_T^l$  & 98.4 & 145.2 & 198.4 & 246.0 & 327.5 & 404.5 & 479.2 & 561.5 \\
$\Delta p^s$ & 1.5 & 0.3 & 0.6 & 1.5 & 2.3 & 2.8 & 3.0 & 2.7 \\
$\Delta p^l$  & 2.8 & 1.9 & 1.1 & 0.6 & 0.1 & 0.0 & 0.6 & 1.7 \\
$-\frac{1}{2}\frac{V_s}{K_T^s}\Delta p_s^2 $ & -1.0 & 0.0 & 0.0 & -0.3 &  -0.6 & -0.6 & -0.5 & -0.4 \\
$-\frac{1}{2}\frac{V_l}{K_T^l}\Delta p_l^2 $ & -3.1 & -1.0 & -0.2  & 0.0 & 0.0 & 0.0 & 0.0 & -0.1 \\
$S^{ls}_{\rm EAM1}$ & 1.04  & 1.01 & 0.98 & 0.96 &0.94 & 0.91 & 0.9 & 0.89  \\
$V^{s}_{\rm EAM1}$ & 11.87  & 11.32 & 10.82 & 10.42 & 10.05 & 9.55 & 9.21 & 8.86 \\
$V^{ls}_{\rm EAM1}$ & 0.636  & 0.537 & 0.452 & 0.404 & 0.364 & 0.318 & 0.295 & 0.279 \\
$\Delta G^{ls}$ & 12.5 (0.9) &  4.9 (1.4) & 0.0 (2.0) & -3.5 (1.4)  & -10.7 (1.8) & -17.2 (1.6) & -27.4 (2.6) & -37.1 (3.0) \\ 
$\delta T_m$ &140 (10) & 56 (16) & 0 (24) & -42 (16) & -132 (22) & -220 (20) & -353 (34)  & -484 (40) \\
$T_m$ & 1637 (10) & 1888 (16) & 2186 (24) & 2461 (16) & 2718 (22) & 3160 (20) & 3522 (34)  &  3857 (40) \\
\end{tabular}
\caption{The various components of the free energy differences between DFT-PBE and EAM1 (see text), the EAM1 melting temperatures and the DFT-PBE melting 
temperatures. Units are GPa for pressure and bulk modulus, 
K for temperature, meV for energy, $k_{\rm B}$ for entropy, and~\AA$^3$/atom for volume.}\label{tab:I}
\end{table}

\begin{table}
\begin{tabular}{ccccccccc}
\hline
$p$ & 80 & 100 \\
$T_m^{\rm EAM2}$ & 3646 & 4049 \\
$\langle \Delta U^{ls}\rangle_{\rm EAM2}$ & -3.7 (0.8) & -9.5 (0.6) \\  
$ -\frac{1}{2k_{\rm B}T}\langle \delta \Delta U_s^2\rangle_{\rm EAM2}$ & -2.5 (0.6) & -4.6 (0.8)  \\
$ -\frac{1}{2k_{\rm B}T}\langle \delta \Delta U_l^2\rangle_{\rm EAM2}$  & -8.2 (1.5)  & -8.0 (1.3) \\
$K_T^s$ & 475.9 & 565.8 &  \\
$K_T^l$  & 430.8 & 577.9 \\
$\Delta p^s$ & 0.7 & 1.5 \\
$\Delta p^l$  & -1.4 & -1.8 \\
$-\frac{1}{2}\frac{V_s}{K_T^s}\Delta p_s^2 $ &  0.0 & -0.1 \\
$-\frac{1}{2}\frac{V_l}{K_T^l}\Delta p_l^2 $ & -0.1  & -0.2 \\
$S^{ls}_{\rm EAM2}$ & 0.89 & 0.89  \\
$V^{s}_{\rm EAM2}$ &  9.21 & 8.87 \\
$V^{ls}_{\rm EAM2}$ & 0.279 & 0.259 \\
$\Delta G^{ls}$ & -9.5 (1.8) & -13.0 (1.6)\\ 
$\delta T_m$ & -124 (24) & -169 (21) \\
$T_m$ & 3522 (24) & 3880 (21)\\
\end{tabular}
\caption{Same as Tab:~\ref{tab:I} but with EAM2 in place of EAM1 (see text)}\label{tab:II}
\end{table}

To compute the DFT-PBE melting curve we applied Eqns.~(\ref{eqn:shift}), (\ref{eqn:dg}), (\ref{eqn:dg1}) as described above. We applied the 
corrections at several values of pressure. This was done by generating long MD simulations with EAM1 at the chosen pressure, both
for solid and liquid, extract from the simulations a large number (typically $>$ 60) of statistically independent configurations, and compute
DFT energies and pressures on these configurations. The simulations have been performed on 256-atom cells, and the DFT 
calculations used $2\times 2 \times 2$ grids of {\bf k}-points, which guarantee convergence of the electronic
free energy to less than 0.05 meV/atom, i.e. leaving a completely negligible error (less than 1 K on the melting temperature). The plane wave cutoff was 337 eV,
which underestimates the pressure by $\simeq 0.4$~GPa. All calculated pressures have been corrected for this small error.
Finally, to test the quality of the PAW potential at high pressure we repeated the calculations at $p=60$~GPa using a PAW potential which
also includes the $3p^6$ electrons in valence. This potential has a core radius of 1.058~\AA, and we used a plane wave cutoff of 460 eV. At
$p=60$~GPa we found $\Delta G^{ls} = 0.015 \pm 0.002$ eV, to be compared to $\Delta G^{ls} = 0.017 \pm 0.002$ eV obtained with the 
small valence PAW, which gives almost exactly the same correction to the melting temperature of EAM1. 

At zero pressure and low temperature nickel has a magnetic moment of  $\simeq 0.62 \mu_B$, which is preserved to at least 200 GPa~\cite{torchio11}.
Above the Curie temperature (628 K at zero pressure) the moments are disordered. Since the melting temperature is well above the Curie temperature, it is likely
that the magnetic moments will be completely disordered in both solid and liquid, and therefore the contribution of magnetic entropy to the free energy difference
between the two phases should cancel out.  Moreover, within the DFT formalism the moments are actually quenched
by the disorder, and since our intention is to provide the DFT melting curve of Ni, all the melting calculations have been performed without including spin-polarisation.   

In Table~\ref{tab:I} we report the free energy difference between the DFT system and EAM1 at several pressures for both solid and liquid,
separated in the various contributions outlined in Eqns.~(\ref{eqn:shift}), (\ref{eqn:dg}), (\ref{eqn:dg1}).  
The correction is positive at low pressures and negative at high pressures, so that the DFT melting slope is lower 
than that of EAM1.  The DFT melting curve is displayed in Fig.~\ref{fig:melt}, where we also plot the recent DAC experimental values of Errandonea 
et al.~\cite{errandonea13}, the SW based experimental values due to Urlin~\cite{urlin66}, 
and the results of previous theoretical calculations by Luo et al.~\cite{luo10}, K\v{o}ci  et al.~\cite{koci06}, and by Weingarten et al.~\cite{weingarten11}.

To cross check the accuracy of Eqn.~(\ref{eqn:shift}) and its range of applicability 
we fitted a second EAM potential in the high pressure region. The parameters of this
second potential were $\epsilon = 3.221\times 10^{-2}$ eV, $a = 3.1773$~\AA, $C=33.0538, n = 8.571, m=3.161$, and we denote as EAM2 the potential with
these parameters. The melting curve of EAM2, together with the DFT corrected melting curve, are also displayed in Fig.~\ref{fig:melt} and reported in Tab.~\ref{tab:II}. 
As expected, the DFT correction is smaller
in the high pressure region, however, the DFT melting curve agrees closely to the DFT melting curve obtained by correcting the melting curve of
 EAM1. This provides a robust validation of Eqn.~(\ref{eqn:shift}). The melting slope of EAM2 is also larger than the DFT one. This behaviour is similar to that observed for
 a similar EAM fitted to reproduce the properties of iron, which also resulted in a steeper melting slope~\cite{belonoshko00} compared to the DFT one~\cite{alfe02}. 
 In Fig.~\ref{fig:melt} we also show the melting curve reported by 
 Luo et al.~\cite{luo10}, and by Weingarten et al~\cite{weingarten11}, both based on empirical potentials. 
 Their melting curves are higher than our DFT melting curve, and higher than the melting curves 
 of both EAM1 and EAM2. On the other hand, the melting curve reported by Ko\v{c}i et al.~\cite{koci06} is quite in good agreement with our ab-initio calculations.
 
As a final cross check to the validity of the calculations, we performed three coexistence simulations directly with DFT 
using a 1000-atom cell. The initial configuration was extracted from a snapshot of an ab-initio
coexistence simulation of aluminium~\cite{alfe03}, with the volume appropriately rescaled so that the pressure was $\simeq$ 8 GPa.
The simulations were performed in the NVE ensemble, using a time step of 3 fs, and the initial value of the internal energy E was set by drawing the initial velocities 
of the atoms from Maxwell-Boltzmann distributions corresponding to initial temperatures of 1700, 2000 and 2300 K. The simulations initiated
with the lowest and the highest temperatures froze and melted, respectively, after  $\simeq 10$ ps, but the third simulation maintained 
coexistence for the whole length of over 30 ps. The average temperature and pressure from this latter simulation are $p = 8.5 \pm 0.1$ GPa and $T= 1896 \pm 10$ K,
which are in excellent agreement with the results obtained by correcting the EAM melting temperature.    

Our DFT-PBE calculated melting temperature of Ni is slightly underestimated by$\sim 90$ K, and the melting slope $dT_m/dp = 30 \pm 2 $ K/GPa is in good agreement with the recent DAC experimental value of 28 K/GPa~\cite{errandonea13}. 
The small underestimate of the zero pressure melting temperature can be understood in terms of the pressure underestimate of DFT-PBE: at the experimental equilibrium volume the DFT-PBE pressure is underestimated by $\sim 3$ GPa, which combined with a melting slope of 30 K/GPa would shift the zero pressure DFT-PBE melting temperature upwards by 90 K, bringing it in perfect agreement with the experimental value.

As pressure increases the calculated melting slope drops less quickly than the DAC experimental one~\cite{errandonea13} and, as a result, at high pressure the DFT-PBE 
melting curve is significantly higher, though very close to old experimental values based on SW~\cite{urlin66}. 

\section{Summary}

In this work we have performed DFT-PBE calculations to compute the melting curve of fcc Ni from 0 to 100 GPa following the approach developed in Ref.~\cite{alfe02b}. 
We find a zero pressure melting temperature of $1637 \pm 10$ K, which is underestimated by about $\sim 90$ K. We argued that this small underestimate can be understood 
in terms of the underestimate of the DFT-PBE pressure at the experimental equilibrium volume.  When we correct for this small error 
the melting temperature comes in perfect agreement with 
the experimental value. The calculated phonon dispersions also agree well with the experimental ones at the 
experimental equilibrium volume.
At high pressure our calculated melting curve deviates from the recent experimental one based on DAC measurements~\cite{errandonea13}, though it appears to agree well
with old experiments based on SW~\cite{urlin66}.

\section*{Acknowledgements}
The work of MP was supported by a NERC grant number
NE/H02462X/1. Calculations were performed on the U.K. national service HECToR.


\begin{thebibliography}{99}

\bibitem{boehler93} R. Boehler, Nature {\bf 363}, 534 (1993).ci
\bibitem{errandonea01} D. Errandonea, B. Schwager, R. Ditz, C. Gessmann, R. Boehler and M. Ross, Phys. Rev. B {\bf 63}, 132104 (2001).
\bibitem{brown86} J. M. Brown and R. G. McQueen, J. Geophys. Res. {\bf 91}, 7485 (1986).
\bibitem{hixson89} R. S. Hixson, D. A. Boness, J. W. Shaner, and J. A. Moriarty, Phys. Rev. Lett. {\bf 62}, 637 (1989).
\bibitem{nguyen04} J. H. Nguyen, and N. C. Holmes, Nature {\bf 427} 339 (2004).
\bibitem{dai09} C. Dai, J. Hu, and H. Tan, J. Appl. Phys. {\bf 106}, 043512 (2009).
\bibitem{santamaria09} D. Santamar\'ia-P\'erez, M. Ross, D. Errandonea, G. D. Mukherjee, M. Mezouar, and R. Boehler,  J. Chem. Phys., {\bf 130}, 124509 (2009).
\bibitem{dewaele10} A. Dewaele, M. Mezouar, N. Guignot and P. Loubeyre, Phys. Rev. Lett. {\bf 104}, 255701 (2010).
\bibitem{jackson13} J. M. Jackson , W. Sturhahn, M. Lerche, J. Zhao, T. S. Toellner, E. E. Alp, S. V. Sinogeikin, J. D. Bass,
C. A. Murphy,  J. K. Wicks, Earth Planet. Sci. Lett. {\bf 362}, 143 (2013).
\bibitem{anzellini13} S. Anzellini, A. Dewaele, M. Mezouar, P. Loubeyre, and  G. Morard, Science {\bf 340}, 464 (2013). See also Y. Fei, Science {\bf 340}, 442 (2013).
\bibitem{laio00} A. Laio, S. Bernard, G. L. Chiarotti, S. Scandolo and E. Tosatti, Science {\bf 287}, 1027 (2000).
\bibitem{belonoshko00} A. B. Belonoshko, R. Ahuja, and B. Johansson, Phys. Rev. Lett. {\bf 84}, 3638 (2000).
\bibitem{luo10} F. Luo, X. R. Chen, L. C. Cai, and G. F. Ji, J. Chem. Eng. Data {\bf 55}, 5149 (2010).
\bibitem{demaske13} B. J. Demaske, V. V. Zhakhovsky, N. A. Inogamov, and I. I. Oleynik, Phys. Rev. B {\bf 87}, 054109 (2013).
\bibitem{weingarten11} N. S. Weingarten and B. M. Rice, J. Phys.: Condens. Matter {\bf 23}, 275701 (2011).
\bibitem{taioli07}  S. Taioli, C. Cazorla, M. J. Gillan and D. Alf\`e, Phys. Rev. B {\bf 75}, 214103 (2007).
\bibitem{alfe02} D. Alf\`e, G. D. Price and M. J. Gillan, Phys. Rev. B, {\bf 65}, 165118 (2002).
\bibitem{alfe09} D. Alf\`e, Phys. Rev. B {\bf 79}, 060101 (2009).
\bibitem{errandonea13} D. Errandonea, Phys. Rev. B {\bf 87}, 054108 (2013).
\bibitem{koci06} L. K\v{o}ci, E. M. Bringa, D. S. Ivanov, J. Hawreliak, J. McNaney, A. Higginbotham, L. V. Zhigilei, A. B. Belonoshko, B. A. Remington, and R. Ahuja
Phys. Rev. B {\bf 74}, 012101 (2006).
\bibitem{mermin65} N. D. Mermin, Phys. Rev. {\bf 137}, A1441 (1965).
\bibitem{perdew96} J. P. Perdew, K. Burke and M. Ernzerhof, Phys. Rev. Lett. {\bf 77}, 3865 (1996).
\bibitem{kresse96} G. Kresse and J. Furthm\"{u}ller, Phys. Rev. B {\bf 54}, 11169 (1996).
\bibitem{blochl94} P. E. Bl\"{o}chl, Phys. Rev. B {\bf 50}, 17953 (1994).
\bibitem{kresse99} G. Kresse and D. Joubert, Phys. Rev. B {\bf 59}, 1758 (1999).
\bibitem{alfe09b} D. Alf\`e, Comput. Phys. Comm. {\bf 180}, 2622 (2009). Program available at http://chianti.geol.ucl.ac.uk/$\sim$dario.
\bibitem{dalcorso13} A. Dal Corso, J. Phys.: Condens. Matter {\bf 25} 145401 (2013).
\bibitem{birgeneau64} R. J. Birgeneau, J. Cordes, G. Dolling, and A. D. B. Woods, Phys. Rev. {\bf 136}, A1359 (1964).
\bibitem{alfe02b} D. Alf\`e, G. D. Price, and M. J. Gillan, J. Chem. Phys., {\bf 116}, 6170, (2002).
\bibitem{frenkel96} D. Frenkel and B. Smit, {\em Understanding Molecular Simulation}, Academic Press, San Diego (1996).
\bibitem{daw83} M. S. Daw and M. I. Baskes, Phys. Rev. Lett. {\bf 50}, 1285 (1983).
\bibitem{sutton90} A. P. Sutton and J. Chen, Phil. Mag. Lett. {\bf 61}, 139 (1990).
\bibitem{hernandez01}E.R. Hern\'andez, J. Chem. Phys. {\bf 115}, 10282 (2001).
\bibitem{urlin66} V. D. Urlin, Sov. Phys. JETP {\bf 22}, 341 (1966).  These results were obtained on the basis of equations of state fitted to the solid and the liquid phases
shock data. 
\bibitem{allen87} M. P. Allen, D. J. Tildesley, {\it Computer Simulation of Liquids}, Oxford Science Publications (1987).
\bibitem{torchio11}R. Torchio, Y. O. Kvashnin, S. Pascarelli, O. Mathon, C. Marini, L. Genovese, P. Bruno, G. Garbarino, A. Dewaele, F. Occelli, and P. Loubeyre, 
Phys. Rev. Lett. {\bf 107}, 237202 (2011).
\bibitem{alfe03} D. Alf\`e, Phys. Rev. B {\bf 68}, 064423 (2003).

\end{thebibliography}

\end{document}